\documentclass[reprint,amsmath,amssymb,notitlepage,citeautoscript,nonbibnotes,nofootinbib]{revtex4-1}

\usepackage{bm,mathrsfs,dcolumn,graphicx,color}
\usepackage{amsmath}
\usepackage{graphicx}
\usepackage[T1]{fontenc}
\usepackage{hyphenat}
\usepackage{setspace}
\usepackage{footnote}
\usepackage[normalem]{ulem}

\definecolor{darkerblue}{rgb}{0,0,0.75}
\definecolor{darkerred}{rgb}{0.8,0,0}
\usepackage[colorlinks=true,citecolor=darkerblue,linkcolor=darkerred]{hyperref}

\begin{document}

\title{Breakdown of the Static Approximation for Free Carrier Screening of Excitons in Monolayer Semiconductors}

\author{Mikhail M. Glazov}
\email{glazov@coherent.ioffe.ru}
\affiliation{Ioffe Institute, 194021 Saint Petersburg, Russian Federation}
\author{Alexey Chernikov}
\email{alexey.chernikov@ur.de}
\affiliation{Department of Physics, University of Regensburg, Regensburg D-93053, Germany}

\begin{abstract}
We address the problem of free carrier screening of exciton states in two-dimensional monolayer semiconductors.
Basic theoretical considerations are presented concerning the applicability of the commonly used static approximation of the screening effect and the implications are discussed. 
We show that the low-frequency models lead to a major overestimation of the free carrier response and are inadequate to describe the screening of strongly bound excitons in monolayer materials.
The presented arguments are consistent with existing high-level many-body theories and transparently illustrate the underlying physics. 
\end{abstract}
\maketitle

\section{Introduction}

The physics of excitons in semiconducting monolayers of transition-metal dichalcogenides (TMDCs) have received a lot of attention in the research community due to the large exciton binding energies, strong light-matter interaction, intriguing spin-valley and exciton transport phenomena\,\cite{Yu2015,Berkelbach2018,Wang2018,Kulig2018}.
With respect to the interaction of the excitons with the solid-state environment, the influence of free carriers is of central importance, both considering fundamental properties and application-driven perspectives.
Moreover, the presence of residual free carriers from non-intentional doping was shown to be ubiquitous in the majority of typically studied samples under most experimental conditions. 
For these reasons, the ensuing phenomena such as the formation of emerging quasi-particles and the corresponding renormalization of the exciton states have been recently under intense experimental and theoretical investigation\,\cite{Wang2018}. 

One of main consequences of finite free carrier concentrations in the material is the dynamical screening of the Coulomb interaction, often described in the Lindhard formalism\,\cite{Stern67a,ando,Haug1989,haug2004quantum}.
It generally includes full frequency and momentum dependence of the screening.
However, it remains rather non-trivial to evaluate excitonic states analytically or even numerically under most circumstances with the dynamic screening taken into account.  
Therefore, a number of approximations was developed in the past to provide more convenient and less demanding description of the screening phenomena within certain limitations.
Among these, the low-frequency approximation, sometimes labeled as the Thomas-Fermi model, proved to be particularly successful to provide both an intuitive and quantitatively accurate treatment of free carrier screening in a variety of semiconducting systems~\cite{ando}.
In TMDC monolayers, the experimental conditions such as carrier concentrations above the degeneracy limit are often comparable to those in more traditional two-dimensional quantum well materials.
Hence, this approach could be naturally considered as the first step to address free carrier screening in these systems as well.

The main distinction of the exciton physics in TMDC monolayer materials from the more traditional quantum well and bulk systems, however, are large binding energies in the 0.5\,eV range which typically exceed by far the Fermi energy of free carriers.
In addition to that, the Coulomb law adopts a non-conventional form, deviating from the reciprocal distance dependence due to the dielectric contrast between the material and its environment~\cite{Rytova1967,Keldysh1979,Cudazzo2011,Berkelbach2013,Cho2018}. 
In the recent literature on the subject, these effects have been taken into account and the free carrier screening has been addressed theoretically using a variety of high-level many-particle calculations~\cite{Steinhoff2014,Schmidt2016,PhysRevB.93.235435,Gao:2016aa,PhysRevB.95.035417,2018arXiv180106217S,Meckbach2018}, resulting in a number of accurate quantitative predictions and descriptions of the experimental findings.
Nevertheless, it remains useful and instructive in this context to explicitly examine the applicability of the low-frequency screening approximation from basic analytical arguments and provide a transparent illustration of the underlying physics.

Here, we address the topic of free carrier screening from this perspective.
We examine the validity of the static model and illustrate the effective breakdown of this approach for monolayer TMDCs due to inapplicability of underlying approximations and in comparison with the existing experimental data.
We outline the main physical reasons and discuss the general implications for the inefficiency of low-frequency approximations for strongly bound excitons in these materials.
We provide basic estimations for the applicability limits and discuss further implications for relevant experimental scenarios. 
Finally, we briefly review the pathways towards calculating a fully dynamic response and discuss the general physics of the effect also in the context of phonon-mediated dielectric screening in monolayer TMDCs.

\section{Free carrier screening of the Coulomb interaction}
\label{sec:scr}

In this section we introduce the general approach to address electron-hole interactions in the presence of free charge carriers.
Here, we take into account the specific screening of the Coulomb interaction in a non-uniform dielectric environment of two-dimensional materials such as TMDC mono- and few-layer systems. 
The excitonic states are further described within the effective mass approximation, where we assume that the energy bands are parabolic and only the direct part of the electron-hole Coulomb interaction is considered, see Ref.~\cite{Wang2018} for details.

The bare (i.e., unscreened by free carriers) Coulomb  potential energy between an electron and a hole is taken in two-dimensional limit of the thin dielectric film model~\cite{Rytova1967,Keldysh1979,Wang2018}
\begin{equation}
\label{vrho}
V(\rho) = -\frac{\pi e^2}{2\varepsilon_s\rho_0} \left[\mathbf H_0\left(\frac{\rho}{\rho_0}\right) - Y_0\left(\frac{\rho}{\rho_0}\right)\right],
\end{equation}
where $\varepsilon_s$ is the average dielectric constant of the surroundings and $\rho_0$ is the screening parameter related to the intrinsic polarizability of the two-dimensional layer~\cite{Cudazzo2011,Berkelbach2013}; $\mathbf H_0$ and $Y_0$ are the Struve and Neumann functions, respectively, and $\rho$ denotes the distance between the electron and the hole in the monolayer plane. 
The Fourier transform of the potential~\eqref{vrho} reads
\begin{equation}
\label{vq}
V_q= \int d\bm \rho V(\rho) e^{-\mathrm i \bm q {\bm \rho}} = -\frac{2\pi e^2}{\varepsilon_s q} \frac{1}{(1+q\rho_0)}.
\end{equation}
The normalization area is set to unity hereafter. Equation~\eqref{vq} differs from the Fourier transform of the Coulomb potential in conventional two-dimensional systems by the presence of an extra factor $(1+q\rho_{0})^{-1}$ which takes into account the dielectric screening in the multilayer structure. 
This factor describes the effect of the spatial dispersion. 
It is related to the fact that the electric field induced by the electron and the hole permeates both the TMDC monolayer and its surrounding medium.
Formally, it corresponds to the explicit dependence of the effective dielectric constant $\varepsilon$ on the wavevector $q$ in the notation 
\begin{equation}
\label{epsilon:def}
V_q=-\frac{2\pi e^2}{\varepsilon{(q)} q},
\end{equation}
with $\varepsilon{(q)}=\varepsilon_s(1+q\rho_{0})$.
In order to retrieve the standard $1/q$ (in the reciprocal space) or $1/\rho$ (in the real space) Coulomb interaction one has to put $\rho_0=0$ in Eqs.~\eqref{vrho} and \eqref{vq}, thus also obtaining a constant $\varepsilon(q)\equiv\varepsilon_s$.

We emphasize that Eqs.~\eqref{vrho} and \eqref{vq} are an approximate representation of the Coulomb interaction in the strict two-dimensional limit.
For realistic structures, they should apply for distances larger than the thickness of the layer.
For excitons, an additional spatial dependence of $\varepsilon_{s}$ and $\rho_0$ can be thus disregarded, at least at a first approximation, if the exciton Bohr radius $a_B$ exceeds by far the lattice constant.
In TMDC monolayer structures, however, $a_B$ exceeds the lattice parameter just by several times.
Hence, while the above approximation was shown to reasonably describe the exciton physics in these materials, the general effects of spatial dispersion on $\varepsilon_s$ and $\rho_0$ should be taken into account for more accurate treatment together with possible deviations from the simple effective mass model, as further discussed in a recent review~\cite{Berkelbach2018} and in the references therein.

Nevertheless, for the purpose of the discussion below, the two-dimensional approximation of the Coulomb interaction is fully sufficient to illustrate our main conclusions, which should be equally applicable for more sophisticated models as well.
More importantly, both $\varepsilon_{s}$ and $\rho_0$ in Eqs.~\eqref{vrho} and \eqref{vq} are generally frequency dependent, i.e., $V_q=V_{q,\omega}$.
This results in the necessity of careful considerations of the proper frequency range to evaluate their precise values.
With respect to that, the arguments presented further below for the case of free carrier screening can be also applied to screening mediated by other electronic and vibronic transitions of the material and its surroundings, as further discussed in the outlook section.

In the following, we consider the system in the presence of free charge carriers, electrons or holes, due to either intentional or unintentional doping. 
Within the random phase approximation the effective screened interaction between an electron and a hole takes the form (also see Refs.~\cite{Stern67a,ando,Haug1989,haug2004quantum} for discussion of various model assumptions and approximations):
\begin{equation}
\label{veff}
V_{q,\omega}^{\rm eff} = \frac{V_q}{1-V_q \Pi(q,\omega)},
\end{equation}
or, alternatively,
\begin{align}
\label{veff:expl}
V_{q,\omega}^{\rm eff} &= -\frac{2\pi e^2}{\varepsilon_s q(1+q\rho_0) + 2\pi e^2\Pi(q,\omega)}\notag
\\&{=-\frac{2\pi e^2}{ \varepsilon_s} \frac{1}{[q+q^2\rho_0 + q_s(q,\omega)]}}.
\end{align}
Here, the polarization function of the free carriers, $\Pi(q,\omega)$, is introduced.
The quantity
\[
2\pi e^2\Pi(q,\omega) \equiv q_s(q,\omega),
\]
can be further associated with the wavevector- and frequency-dependent screening wavevector $q_s(q,\omega)$.
In the linear screening regime the polarization function can be expressed in the form
\begin{equation}
\label{Pi}
\Pi(q,\omega) = g \sum_{\bm p } \frac{f(\bm p+\bm q) - f(\bm p)}{E_{\bm p+ \bm q} - E_{\bm p} - \hbar\omega - \mathrm i \delta }.
\end{equation}
Here, $E_{\bm p} = p^2/(2m)$ is the free carrier dispersion, $m$ is the effective mass, $g$ is the degeneracy factor accounting for the number of the occupied spin and valley states.
For TMDCs, $g=2$ if only bottom conduction subbands  are filled by the electrons (or topmost valence subbands are filled by the holes) or $g=4$ if both spin subbands are filled in each valley of the energy spectrum. In the latter case, finite energy separations between the bands need to be further taken into account to evaluate the expression in Eq.\,\eqref{Pi}.
The charge carrier Fermi-Dirac distribution function $f(\bm p)$ is given by 
\begin{equation}
f(\bm p) = \frac{1}{1+\exp{\left(\frac{E_{\bm p} -\mu}{k_B T}\right)}},
\end{equation}
where $T$ is the temperature, $\mu$ is the chemical potential, and the infinitesimal parameter $\delta\to+0$ ensures the causality. 
In general terms, the screening is due to the fact that the electric field produced by the electron-hole pair, for example, affects the free charge carriers and perturbs their distribution in the real and momentum space. 
As a result of this redistribution, an imbalance of the electric neutrality occurs and the electron-hole interaction becomes screened at certain distances.

Strictly speaking, Eqs.~\eqref{veff} and \eqref{Pi} are valid if the Coulomb interaction between free charge carriers is relatively weak compared to their kinetic energy.
Formally, the carriers' Fermi energy $E_F =  2\pi n\hbar^2/(mg)$ should exceed the characteristic Coulomb interaction energy of the charge carriers, $E_C  =e^2\sqrt{n}/\varepsilon_{s}$. 
Accordingly, the density parameter $r_s = (\sqrt{\pi n} a_B)^{-1}$, with $a_B$ being the exciton Bohr radius, should be small, $r_s \ll 1$.
As a consequence, this seems to apparently require experimental conditions at the exciton ionization threshold or Mott transition.  
However, for realistic electron or hole densities across a number of instances where the model was successfully applied the parameter $r_s\gtrsim 1$~\cite{ando,giuliani2005quantum}.
As an example, for monolayer TMDCs, taking electron density $n=3\times 10^{12}$~cm$^{-2}$ ($r_s = 3.3$ for $a_B=1$\,nm), $g=2$, $m=0.5m_0$ with $m_0$ being the free electron mass, and $\varepsilon_s=4$ we have the Fermi energy $E_F\approx 14$~meV and the Coulomb energy $E_C \approx 62$~meV.
Indeed, a more detailed analysis demonstrates~\cite{PhysRevB.39.5005,giuliani2005quantum} that the interactions between the resident charge carriers dominate the state of the electron/hole gas at much lower densities, where $r_s \sim 30$. Thus, the considerations presented in the following should apply for the typical density regime of free carrier densities in TMDC monolayers between 10$^{11}$ and 10$^{13}$\,cm$^{-2}$, roughly corresponding to $r_s$ between 18 and 2, respectively.

Generally, in the presence of free charge carriers, the polarization function in Eq.~\eqref{Pi} (or the screening wavevector $q_s$ in Eq.~\eqref{veff:expl}) depends on both the frequency $\omega$ and the wavevector $q$. 
Essentially, the electron or hole gas should accommodate and rearrange itself with respect to the ``external'' potential, e.g., introduced by the electron and hole forming an exciton. 
The process is naturally delayed in time and also depends on the spatial scale, i.e., the wavelength, of the perturbation (or the value of $q^{-1}$), potentially leading to vastly different screening efficiencies under different circumstances. 
With respect to that, the basic parameters of the charge carrier gas such as the ``Fermi frequency'' $\omega_F =E_F/\hbar$ and the Fermi wavevector $k_F = \sqrt{2mE_F/\hbar^2}$ set natural bounds on the screening response.
More specifically, \textit{the perturbations with frequencies $\omega \gtrsim \omega_F$ and/or wavevectors $q\gtrsim k_F$ cannot be screened effectively} by the free carriers. 
Such perturbations change in time and/or space too rapidly: the carriers are unable either to follow these fast enough or to arrange themselves on sufficiently small spatial scales.
While trivially following from basic physics and from Eq.~\eqref{Pi}, this particular point is both of general importance for the evaluation of the screening efficiencies and is highly relevant specifically for the case of TMDC monolayers in particular.
We also note that additional restrictions on the screening efficiency are imposed due to excitation of plasmons in the free carrier gas, see Ref.~\cite{Chaplik:1985aa} for review. 

However, the explicit treatment of both frequency and spatial dependence of the polarization function in Eq.~\eqref{Pi} is usually extremely challenging and thus often requires additional approximations.
Considering spatial dependence, taking the long wavelength limit $\bm q\to 0$ directly yields the classical Drude response~\cite{haug2004quantum}, typically used to describe screening in simple metals.
With respect to frequency, however, the \textit{static screening approach} is, as a rule, applied in order to calculate the exciton properties in the presence of free carriers in more conventional two-dimensional semiconductor systems, such as GaAs quantum wells.~\cite{doi:10.1063/1.335993,pikus_eng_92}. 
In this limiting case, the polarizability is replaced by its value at $\omega\to0$. 

\begin{figure}[t]
\includegraphics[width=8.2 cm]{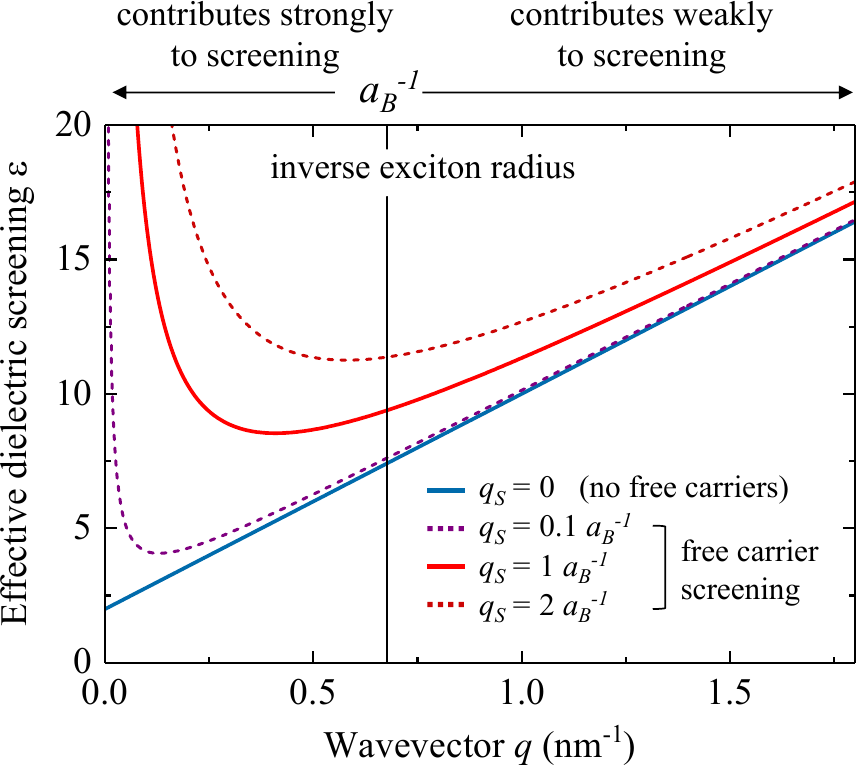}
\caption{Illustration of the effective dielectric screening $\varepsilon=\varepsilon_s(1+q\rho_0+q_s/q)$ from the static approximation in Eq.\,\eqref{veff:static}.
It is defined according to Eq.~\eqref{epsilon:def} and shown both in the presence or absence of free charge carriers.
The parameters are chosen to roughly match the typical case of a TMDC monolayer on glass substrates: $\rho_0=4$\,nm and $\varepsilon_s=2$.
The corresponding reciprocal exciton Bohr radius in the zero-density case of $a_B^{-1}$\,=\,(1.5\,nm)$^{-1}$ is obtained assuming equal electron and hole effective masses of $m=0.5\,m_0$ and indicated by the vertical line.
The screening wavevector values are chosen to represent the weak ($q_s=0.1\,a_B^{-1}$) and strong ($q_s={a_B}^{-1}$, $q_s=2\,a_B^{-1}$) screening regimes.
The arrows on top further indicate the ranges of the wavevector values $q$ roughly contributing either strongly or weakly to the screening of the exciton with the radius $a_B$.
}
\label{fig:1}
\end{figure}

Then, the sum in Eq.~\eqref{Pi} can be readily calculated with the result $\Pi = gmf(0)/(2\pi\hbar^{2})$. 
Hence, in the case of the static screening one obtains the effective Coulomb potential from Eq.\,\eqref{vq} modified to
\begin{equation}
\label{veff:static}
V_{q, {\omega\to0}}^{\rm eff} = -\frac{2\pi e^2}{ \varepsilon_s} \frac{1}{(q+q^2\rho_0 + q_s)}.
\end{equation}
It formally resembles Eq.\,\eqref{veff:expl}, albeit with a \textit{frequency-independent} screening wavevector $q_s$ given by
\begin{equation}
\label{qs}
q_s = \frac{gme^2}{\hbar^{2} \varepsilon_s} f(0) = \frac{gme^2}{\hbar^{2}  \varepsilon_s}\left[1- \exp{\left(-\frac{2\pi \hbar^2 n_e}{gm k_B T}\right)} \right].
\end{equation}
It thus generally depends on both carrier density and temperature.
At low temperatures or high carrier densities, however, where 
\[
\frac{2\pi \hbar^2 n_e}{gm} \gg k_B T,
\]
the exponent is vanishing, the screening wavevector can be further approximated by a constant value
\begin{equation}
\label{qs:low}
q_s = \frac{gme^2}{\hbar^{2}  \varepsilon_s}.
\end{equation}

The resulting dependence of the effective dielectric susceptibility $\varepsilon\,{=\varepsilon_s\,(1+q\rho_0+q_s/q)}$ introduced according to Eqs.~\eqref{epsilon:def} and \eqref{veff:static} on the wavevector $q$ in the presence of free charge carriers is illustrated in Fig.\,\ref{fig:1}.
While the axes can be normalized to an arbitrary inverse exciton radius and static screening constant, we choose the parameters roughly representing the typical case of a TMDC monolayer on glass substrates to obtain quantitative results in reasonable agreement with experiments: $\rho_0=4$\,nm and $\varepsilon_s=2$.
The corresponding reciprocal exciton Bohr radius in the zero-density case of $a_B^{-1}$\,=\,(1.5\,nm)$^{-1}$ is calculated assuming equal electron and hole effective masses of $m=0.5\,m_0$, as discussed further below.
The screening wavevector values are chosen to represent the weak ($q_s=0.1\,a_B^{-1}$) and strong ($q_s={a_B}^{-1}$, $q_s=2\,a_B^{-1}$) screening regimes as well as the case in the absence of free carriers ($q_s=0$).

As long as the screening wavevector $q_s$ is much smaller than $a_B^{-1}$, the Coulomb interaction is efficiently screened only at a relatively long range compared with the typical distances between the electron and the hole in an exciton.
In this regime, the properties of the exciton ground state are barely affected by the presence of free carriers.
When $q_s$ increases towards $a_B^{-1}$, however, the screening becomes increasingly efficient. 
For almost complete screening of the exciton, the order of magnitude estimate for the screening wavevector is typically taken as $q_s \sim a_B^{-1}$.
Then, in the static screening approach the Coulomb attraction between the electron and hole at the relevant distances $\rho \sim a_B$ is strongly reduced in the presence of degenerate charge carriers. 
As a consequence, the exciton binding energy should drastically decrease with increasing in the doping density with $q_s$ approaching $a_B^{-1}$ and, as soon as the electron or hole gas becomes degenerate, the excitons are expected to almost vanish. 

We also note that the above arguments generally apply for the excited states of the exciton as well by considering their respective inverse radii instead of $a_B^{-1}$.
Since the wavefunctions of these states spread of larger distances in the real space, corresponding to smaller $q$ in the reciprocal space, in comparison to the ground state, they are generally more susceptible to screening by free carriers.
In close analogy, with respect to the frequency dependence discussed below, similar considerations apply with respect to the binding energies of the excited states being much smaller than that of the ground state.

\section{Static screening of excitons in 2D TMDC: Case study}
\label{sec:stat}

In this section we examine the consequences of a finite screening wavevector $q_s$ on the binding energies and radii of the exciton states for parameters typical for TMDC monolayers.
We analyze the differences associated with the modified thin-film Coulomb potential in contrast to the more traditional interaction with the reciprocal distance dependence.
Then, we discuss the quantitative predictions of the static screening approximation and compare these with the experimental observations from the literature.

In the presence of free carrier screening, the excitonic states are found by solving the Schr\"odinger (or Wannier) equation in the momentum space:
\begin{equation}
\label{schroed}
\frac{\hbar^2k^2}{2\mu} \psi_{\bm k} + \sum_{\bm q} V_{q}^{\rm eff} \psi_{\bm k-\bm q} = -E_B \psi_{\bm k}.
\end{equation}
Here, $\mu$ is the reduced mass of the electron-hole pair, $\psi_{\bm k}$ is the Fourier transform of the exciton relative motion wavefunction and $E_B$ is the exciton binding energy.
The binding energy is defined, as usual, with respect to the continuum of free-particle states, i.e., the band gap $E_g$ (which can be in principle also renormalized by the presence of free carriers~\cite{Wang2018,Klingshirn2007}).
The potential $V_{\bm q}^{\rm eff}$ is given by Eq.~\eqref{veff:static} and the Eq.\,\eqref{schroed} is solved numerically using variational approach with the two-dimensional hydrogenic trial function.
We note that while there are more complex and accurate solutions to the numerical problem, the variational approach provides reasonable quantitative results and is fully sufficient to illustrate the physics discussed below.

\begin{figure}[t]
\includegraphics[width=8.4 cm]{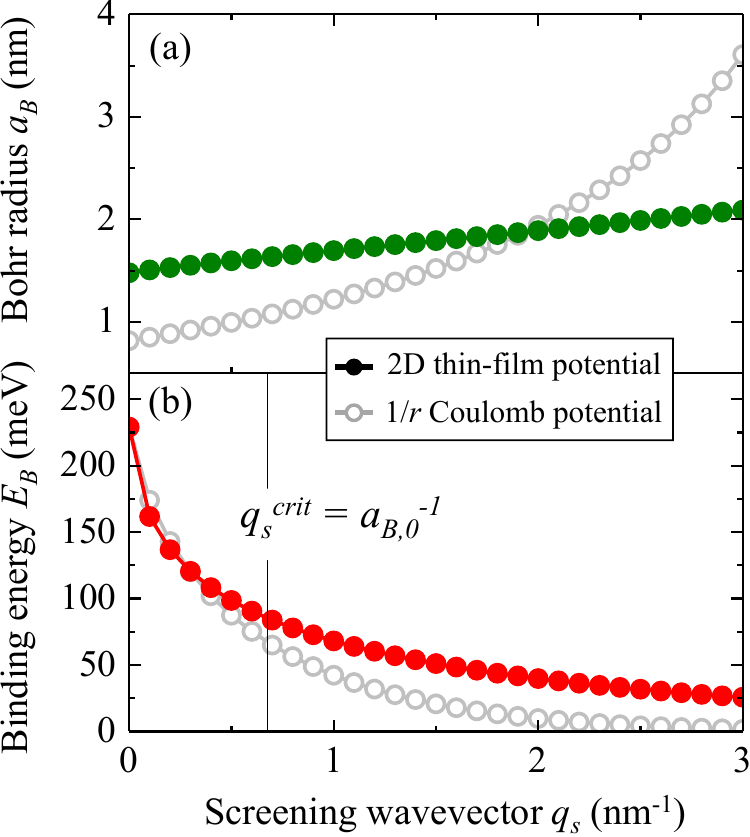}
\caption{Results of the variational solution of Eq.\,\eqref{schroed} with $\mu=0.25\,m_0$ for the 2D thin-film potential ($\rho_0=4$\,nm, $\varepsilon_s=2$) and the traditional reciprocal distance dependence of the Coulomb interaction ($\rho_0=0$\,nm, $\varepsilon_s=7.7$).
(a) The top and (b) the bottom panels show the exciton Bohr radius $a_B$ and the binding energy of the ground states $E_B$, respectively.
The critical value of the screening wavevector $q_s^{crit}$ matching the inverse Bohr radius $a_{B,0}^{-1}$ calculated for the thin film potential in the absence of screening is indicated by the vertical line.}
\label{fig:2}
\end{figure}

Equation~\eqref{schroed} generally applies for any potential of the form of Eq.~\eqref{veff:static}.
For the free carrier screening, it corresponds to the case of static approximation as discussed in Sec.~\ref{sec:stat} and in the following Sec.~\ref{sec:dyn}.  It is, moreover, valid under the condition of $k_F \ll a_B$. 
Otherwise, the state filling effects should be taken into account, e.g., by enforcing $\psi_{\bm k}$ to be zero for $k\leqslant k_F$~\cite{pikus_eng_92}.
We note that even in this regime, additional many-body effects such as trion formation and appearance of Mahan singularity along with more complex correlations in the Fermi sea of resident charge carriers may arise~\cite{Wang2018,Mak:2013lh,PSSB:PSSB343,suris:correlation,PhysRevB.95.035417,Sidler:2016aa}. 
The state filling effects, however, while being of general importance to describe the full response of the system at moderate and high carrier densities, are beyond the scope of the present paper.
Also, the requirement $k_F \ll a_B$ or, alternatively, $E_F \ll E_B$ holds in most practical cases in TMDC crystals under typical experimental conditions. 

For the quantitative analysis, we fix the system parameters to roughly representative values for TMDC monolayers: $\mu=0.25\,m_0$, $\rho_0=4$\,nm, $\varepsilon_s=2$.
The latter approximately corresponds to a monolayer supported by a glass substrate.
In addition, we also consider the case of the traditional hydrogenic Coulomb potential with the $1/\rho$ distance (or $1/q$ wavevector) dependence.
For that, we set $\rho_0=0$\,nm and adjust $\varepsilon_s$ to 7.7 to match the exciton binding energy obtained with the thin-film potential without free carrier screening for the parameters chosen above.
In both cases, the band degeneracy factor $g$ is set to 2, corresponding to the filling of the lowest conduction bands (or highest valence bands) at $K^{+}$ and $K^{-}$ valleys.

The resulting dependence of the exciton radius and binding energy on the screening wavevector $q_s$ is presented in the top (a) and bottom (b) panels of Fig.\,\ref{fig:2}, respectively.
We note that at this stage no specific relation between $q_s$ and the free carrier density $n$ needs to be considered and the results depend on $q_s$ alone. 
As the screening wavevector increases, corresponding to decreasing spatial distance of efficiently screened Coulomb interaction discussed in Fig.\,\ref{fig:1}, the exciton Bohr radius becomes larger and the binding energy decreases. 
At the aforementioned critical value $q_s^{crit}$ equal to the inverse exciton Bohr radius at zero-screening $a_{B,0}^{-1}$ the binding energy decrease to about one-third of its initial value.
For even larger $q_s$, it rapidly converges towards zero, reflecting a severely weakened electron-hole interaction in the exciton.

It is noteworthy that both modified and bare Coulomb potentials lead to a rather similar dependence of the binding energy on the screening wavevector.
In contrast to that, the change in the exciton radius for the hydrogenic 1/${\rho}$ interaction is much more pronounced compared to the thin-film case.
As a consequence, while the binding energy decreases rapidly, the oscillator strength of the optical resonance, roughly proportional to $a_{B}^{-2}$ should change more slowly for the non-hydrogenic modified potential in a monolayer.
Nevertheless, even in this case, the radius will eventually increase further for higher $q_s$ and the strength of the exciton transition will finally approach values close to zero.

\begin{figure}[t]
\includegraphics[width=8.4 cm]{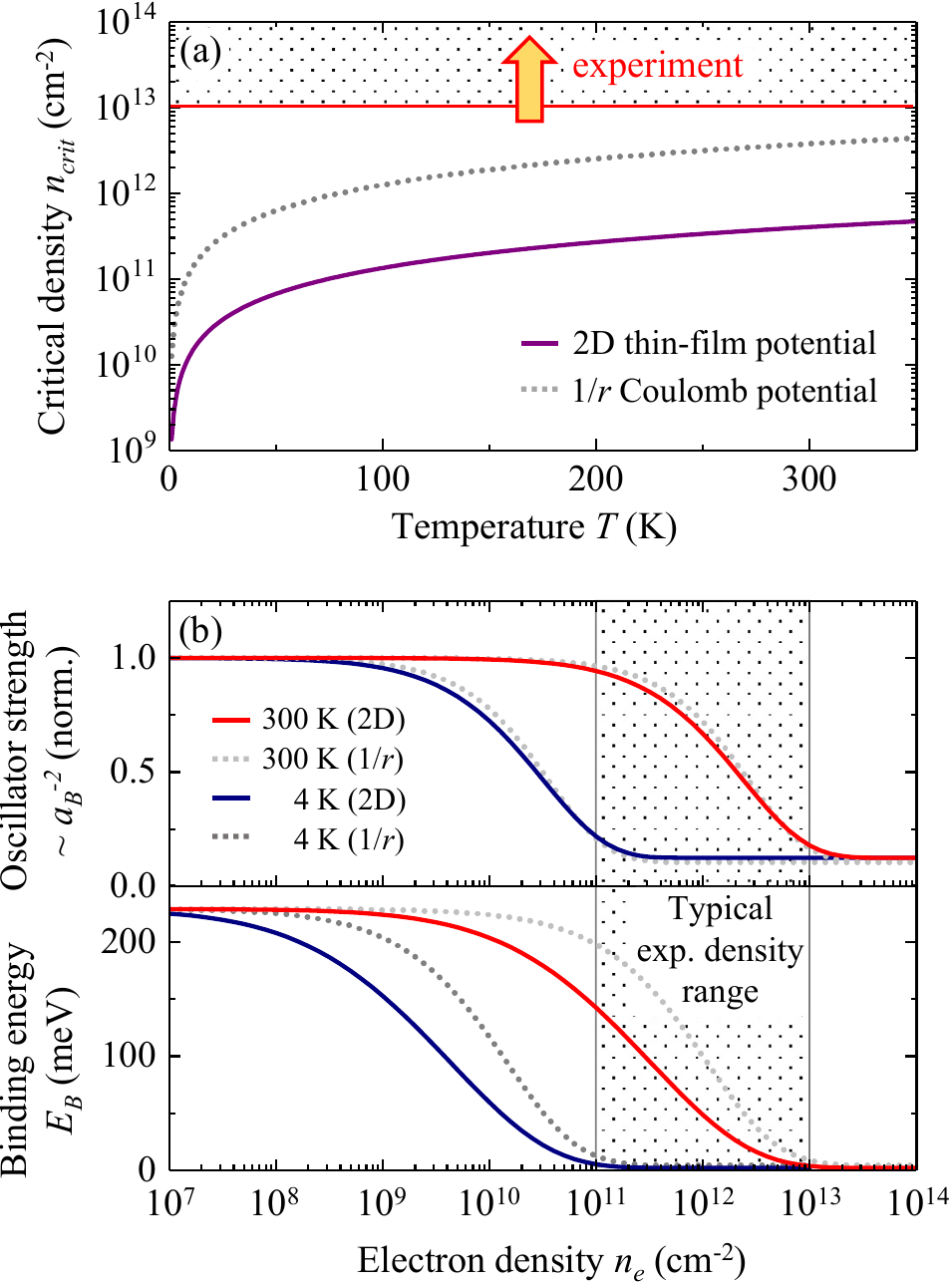}
\caption{(a) Critical free carrier carrier density $n_{crit}$ corresponding to the condition ~\eqref{critical} for efficient screening of excitons in TMDC monolayers as a function of temperature in static approximation for two different Coulomb potentials.
The corresponding density range of exciton dissociation consistent with experimental observations is indicated by the shaded area.
(b) Normalized inverse squared exciton radius, proportional to the oscillator strength (upper panel), and the exciton binding energy (lower panel) as function of the free electron density calculated in the static screening approximation for cryogenic and room temperatures.
Typical accessible density range in the experiment for both intentionally and unintentionally doped samples is indicated by the shaded area.}
\label{fig:3}
\end{figure}

Next, we explicitly evaluate the dependence of the screening wavevector $q_s$ on the free carrier density in the static approximation according to Eq.\,\eqref{qs}.
First, the critical density $n_{crit}$ corresponding the previously discussed condition 
\begin{equation}
\label{critical}
q_s^{crit}\equiv q_s(n_{crit})=a_{B,0}^{-1},
\end{equation}
is presented in Fig.\,\ref{fig:3}\,(a) as function of temperature.
Since it is proportional to the value of the Fermi distribution at zero, it exhibits a strong temperature dependence and varies by more than two orders of magnitude between cryogenic and room temperatures.
Overall, the critical densities are significantly lower in the case of the modified thin-film Coulomb potential in comparison to the hydrogenic one, ranging from $10^{9}$ to several $10^{11}$\,cm$^{-2}$.
Nevertheless, both models predict highly efficient screening of excitons already for rather low carrier densities at all temperatures.
As also discussed further below, this does not seem to be consistent with experimental reports in TMDC monolayers\,\cite{Chernikov2015c,Chernikov2015b,Wang2018} and recent high-level many-body calculations\,\cite{2018arXiv180106217S} where the exciton dissociation is typically found at much higher carrier densities on the order of $10^{13}-10^{14}$\,cm$^{-2}$, as indicated in Fig.\,\ref{fig:2}\,(a).
At cryogenic temperatures, in particular, the discrepancy is as high as at least two orders of magnitude.

The explicit dependence of the exciton binding energy and the normalized inverse squared exciton radius $a_B^{-2}$, proportional to the oscillator strength, on the free electron density is presented in the bottom and top panels of Fig.\,\ref{fig:3}\,(b).
It is shown in the temperature range between 4 and 300\,K and for the two forms of the Coulomb potential.
The free carrier densities typically accessed in the experiments for the samples with both intentional and unintentional doping are roughly indicated by the shaded area\,\cite{Mak2012b,Ross2013,Jones2013,Chernikov2015b,Courtade2017,Wang2018}.
In this range, the static approximation of the free carrier screening predicts strongly reduced binding energies and oscillator strengths as compared to the pristine case.
Moreover, as the system is cooled down to 4\,K, the excitons are expected to be almost dissociated at these densities and the corresponding resonance in optical spectra should disappear.
In addition, both the exciton oscillator strength and binding energy should strongly change with temperature, essentially following the dependence presented in Fig.\,\ref{fig:3}\,(a).

In the experiments on TMDC monolayers, however, neither of these effects are observed.
The total oscillator strength of the exciton resonance, proportional to the area of the resonance in optical absorption or reflectance, is barely affected by free carriers up to few 10$^{12}$\,cm$^{-2}$ and is only partially redistributed between the neutral and charged exciton transitions\,\cite{Mak2012b,Ross2013,Chernikov2015b}.
The oscillator strength is also almost temperature independent in both intentionally and unintentionally doped samples with the usual densities below 10$^{13}$\,cm$^{-2}$\,\cite{Arora2015,Arora2015a,Chernikov2015b}.
Finally, the binding energy measured in unintentionally doped samples with densities typically on the order of several 10$^{12}$\,cm$^{-2}$ in most cases and often at low temperatures are found to be several hundreds of meV\,\cite{Wang2018} rather than being vanishingly small as shown in Fig.\,\ref{fig:3}\,(b).

This means that the \textit{static approximation}, $\omega\to0$, in Eq.~\eqref{veff} (see also Eq.~\eqref{veff:static}) of the free carrier screening \textit{severely overestimates the screening efficiency} and is inconsistent with the experimental observations.
This conclusion is also largely independent from the specific choice of the Coulomb potential, as shown above.
Moreover, for the properly modified electron-hole interaction in monolayer materials, the discrepancies are even larger than for the hydrogenic case.
From theoretical point of view, the breakdown of the static approach can be rationalized by self-consistently examining the frequency range necessary to efficiently screen the excitons in 2D TMDCs.
As obtained by the solution of Eq.\,\eqref{schroed} for representative TMDC monolayer parameters, binding energies of the exciton ground state in the range of 100's of meV are significantly higher than the Fermi energies under typical conditions by at least one order of magnitude. 
Thus, the free charge carriers are essentially too slow and unable to follow the movement of the electron and hole in the exciton fast enough and rearrange themselves accordingly, leading to inefficient screening.

This is in strong contrast to more traditional 2D systems, such as the quantum wells, with exciton binding energies orders of magnitude smaller than in the TMDC monolayers.
Hence, only if the binding energy is reduced to several 10's of meV, approaching the range of Fermi energies corresponding to carrier densities below $10^{13}$\,cm$^{-2}$, the free carrier screening becomes efficient.
Overall, the presence of free charge carriers at these densities should certainly affect the exciton properties of monolayer TMDCs, as also observed in the experiments, albeit not as drastically as predicted by the static approach.

\section{Dynamic screening of excitonic states}
\label{sec:dyn}

In the previous section we have shown the inadequacy of the static approximations to address free carrier screening in semiconducting monolayer materials.
As a consequence, dynamic approaches taking full frequency dependence of the screening into account are required.
In this last section, we would like to review general considerations towards addressing the problem of free carrier screening of excitons in systems with large binding energies, such as 2D TMDCs.
We note in advance that theoretical calculations of dynamic screening are very challenging, so that providing detailed and accurate solutions remains far beyond the scope of this work.
Our intent is thus to outline the scope of the problem and illustrate possible directions towards the solution.

\begin{figure}[t]
\includegraphics[width=\linewidth]{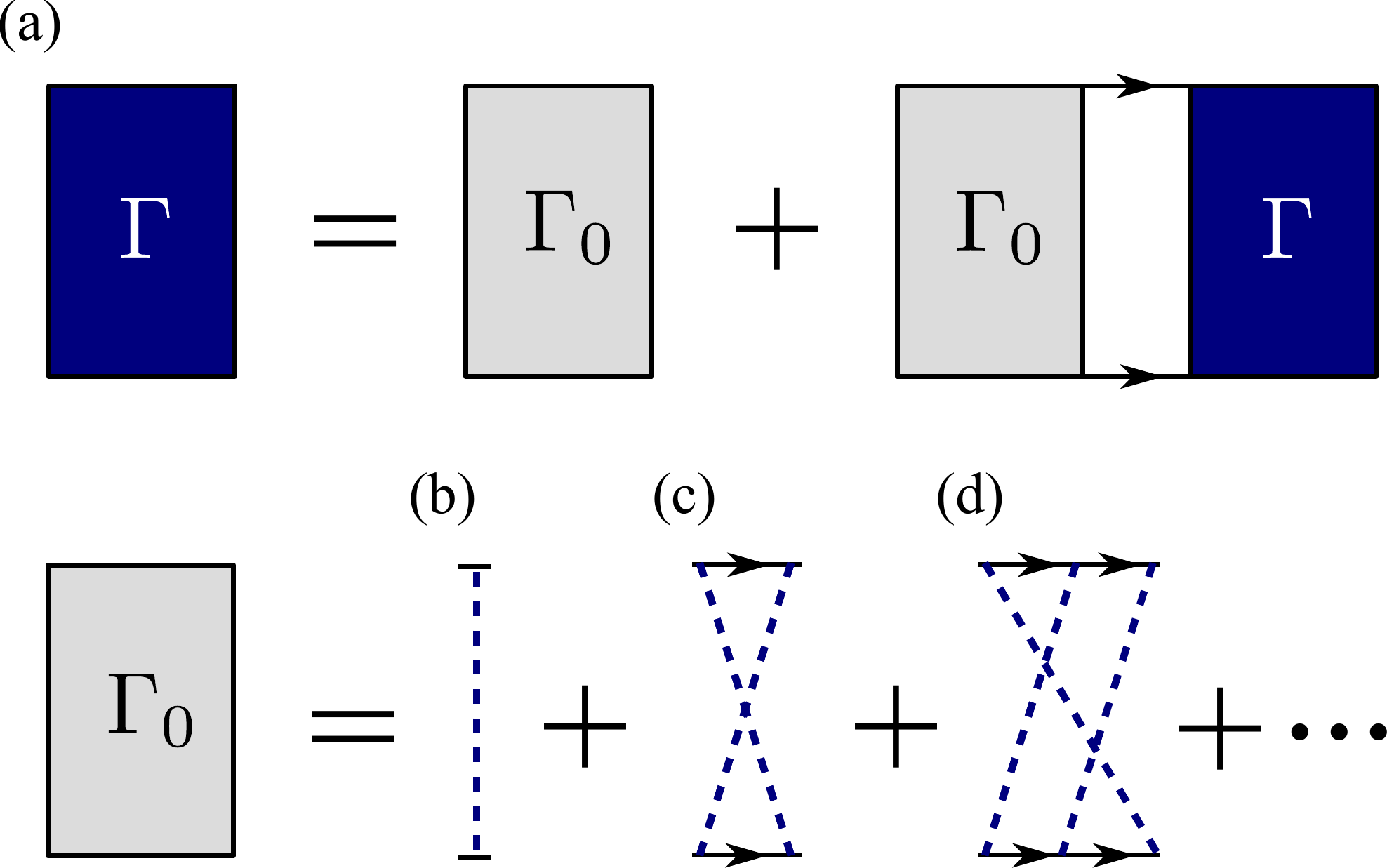}
\caption{(a) Schematic representation of the Bethe-Salpeter equation. Solid lines with arrows show electron and hole Greens functions, dark  rectangle is the two-particle interaction vortex, light rectangle is the bare (irreducible) two-particle interaction vortex. (b\,--\,d) Diagrams representing the irreducible vortex. Dashed line is the screened Coulomb interaction, Eq.~\eqref{veff}. Dots denote omitted higher order in $V^{\rm eff}_{\omega,q}$ terms.}\label{fig:BS}
\end{figure}

Generally, the account for the dynamic screening of excitonic states, i.e., of the retardation in the electron-hole interaction represents an extremely complex problem. In this regard, an analogy with a problem of \emph{positronium}, a bound state of an electron an a positron, can be drawn~\cite{ll4_eng,akhiezer}. The rigorous approach in both cases is based on the solution of the Bethe-Salpeter equation for the two-particle Greens function. Graphically, this equation is presented in Fig.~\ref{fig:BS}. It involves the two-particle interaction vortex $\Gamma$, which is determined by the infinite series of the diagrams involving different powers of the Coulomb interaction. The latter can be formally summed up resulting in the integral equation shown in Fig.~\ref{fig:BS}(a), whose kernel is expressed via the sum of the irreducible contributions. Several lowest order ones are depicted in the panels (b) (first order), (c) (second order), and (d) (third order) of Fig.~\ref{fig:BS}.

If the occupation of the bands can be neglected and, moreover, if the screening is static, in other words, if $V^{\rm eff}_q$ depends on the transferred momentum only, the contributions with self-crossings [Fig.~\ref{fig:BS}(c,d) and omitted higher order terms] vanish, which can be readily checked by using the space-time representation for the Greens functions. In this situation the irreducible vortex is given simply by the interaction potential reducing the problem to the discussed above in Sec.~\ref{sec:stat}. The account for the retardation, i.e., for the explicit dependence of $V^{\rm eff}_{\omega,q}$ on the frequency $\omega$, makes the contributions (c), (d), \ldots non-zero. Similarly, the presence of the resident electrons or holes in the bands provides contribution to the diagrams in Fig.~\ref{fig:BS}(c,d). This makes the problem of calculation of exciton in the presence of free charge carriers barely solvable and requires making further approximations.

As a starting point one usually omits the self-crossing diagrams and replaces $\Gamma_0$ by the dynamically screened Coulomb potential, Fig.~\ref{fig:BS}(b). Technically, it is convenient to 
use the reference frame where the exciton is at rest and 
introduce the function~\cite{ll4_eng} $\chi(\bm k, \Omega, \delta) = G^{e}(\bm k, \Omega/2 + \delta)\Gamma G^{h}(\bm k, {\Omega}/{2} - \delta)$, which differs from the two-particle vortex by the presence of additional Greens functions of the electron and the hole:
\begin{subequations}
\label{Greens}
\begin{align}
G^{e}(\bm k, \Omega) = \frac{1}{E^{e}_{\bm k} - \hbar \Omega + \mathrm i 0},\\ 
 G^{h}(\bm k, \Omega) = \frac{1}{E^{h}_{\bm k} -  \hbar \Omega + \mathrm i 0},
 \end{align}
\end{subequations}
where $E^{e}_{\bm k}$, $E^{h}_{\bm k}$ are electron and hole dispersions, $\hbar\Omega$ and $\hbar\delta$ are the energy variables. In order to find the binding energies it is sufficient to determine the poles of $\Gamma$ or of $\chi$ as a function of $\Omega$, i.e., the energies $\hbar\Omega$ where $\Gamma$ becomes infinite. In this case, the first term in Fig.~\ref{fig:BS}(a), can be omitted, and the Bethe-Salpeter equation for the exciton at rest (center of mass momentum is zero) takes the form~\cite{ll4_eng,PhysRev.144.708,makarov:68,PSSB:PSSB2220480218,PSSB:PSSB2220850219,averkiev_monakhov,2018arXiv180106217S}
\begin{multline}
\label{BS1:1} 
\left(E_g + \frac{\hbar^2k^2}{2\mu} - \hbar \Omega\right)\chi(\bm k, \Omega, \delta) =\\ 
\mathrm i \left[G^{e}\left(\bm k, \frac{\Omega}{2} + \delta\right) + G^{h}\left(\bm k, \frac{\Omega}{2} - \delta\right)\right] \times \\
\sum_{\bm q} \int \frac{d\delta'}{2\pi}  V_{q,\delta-\delta'}^{\rm eff} \chi(\bm k - \bm q, \Omega, \delta').
\end{multline}
This is so-called screened ladder approximation. Here we assumed for simplicity that electron and hole have the same effective mass, $\bm k$ is the relative motion wavevector. In Eq.~\eqref{BS1:1} the occupancy of the conduction band can be neglected, the energies  $E^{e}_{\bm k}$, $E^{h}_{\bm k}$ include, in general, contributions due to the resident carriers interaction between themselves.

In the case of the static screening, where $V_{q,\delta-\delta'}^{\rm eff}$ does not depend on $\delta - \delta'$, the dependence on the $\delta$-variable can be excluded by integrating both sides of Eq.~\eqref{BS1:1} over $\delta$. In which case for the function $\psi_{\bm k} = \int \frac{d\delta}{2\pi} \chi(\bm k, \Omega, \delta)$,
we recover Schr\"odinger Eq.~\eqref{schroed} with the statically screened potential and the binding energy $E_B = E_g - \hbar \Omega$.
 
The frequency independent form of the interaction can be applied if characteristic frequencies $\omega_*$ (where $V_{\bm q,\omega_*}^{\rm eff}$ either has resonances or changes very strongly) are very different from the frequency of electron-hole relative motion in the exciton $\omega_B = E_B/\hbar$.
It directly follows from the characteristic argument of the potential in Eq.~\eqref{BS1:1} being $|\delta - \delta'| \sim E_g - \hbar \Omega =E_B$. 
Particularly, if $\omega_* \gg \omega_B$ the typical frequencies transferred in the course of electron-hole interaction in Eq.~\eqref{BS1:1} are small as compared with $\omega_*$ and the potential $V^{\rm eff}_{\bm q,\delta - \delta'}$ can be replaced by its zero-frequency value, $V^{\rm eff}_{q,0}$. By contrast, if $\omega_* \ll \omega_B$, the high-frequency asymptotics, $V^{\rm eff}_{q,\omega\to \infty}$, has to be used.

In the situation where the screening is provided by the resident charge carriers, the typical frequencies where the $V^{\rm eff}_{\bm q,\omega}$ varies strongly are, as discussed above, related to the Fermi frequency $E_F/\hbar$. For electron or hole densities in TMDC monolayers up to $10^{12}\ldots 10^{13}$~cm$^{-2}$ the Fermi energy is much smaller than the exciton binding energies $\sim 200\ldots 500$~meV and the high frequency asymptotics for the interaction potential should be used. In this case, essentially, $\Pi(q,\omega)\to 0$ in Eq.~\eqref{Pi} and $V^{\rm eff}_{q,\omega} \equiv V_q$, i.e., the interaction is not screened. This is in stark contrast with the more traditional III-V or II-VI semiconductor based quantum wells where exciton binding energies are often in the range $\sim 10$~meV, thus, being comparable or smaller than characteristic Fermi energy of the resident electron gas in the presence of intentional doping. In this conventional situation one can use the static screening approximation and, moreover, account for a finite occupancy of the conduction or valence bands.
Nevertheless, we note that the presented arguments are not limited to specific material systems and should generally apply as soon as exciton binding energy becomes much larger than the Fermi energy of the free carriers.

In order to account for the dynamical screening in Eq.~\eqref{BS1:1} various approximations are invoked. The simplest approach is to consider an effectively static screening with the replacement
\begin{equation}
\label{repl}
V^{\rm eff}_{q,\delta - \delta'} \to V^{\rm eff}_{q,E_B/\hbar},
\end{equation}
and solving the generalized Schr\"odinger equation~\eqref{schroed} with $V^{\rm eff}_q$ dependent on the eigenvalue $E_B$~\cite{makarov:68,Gao:2016aa}, see also Ref.~\cite{2018arXiv180106217S}. In the more advanced approach of Ref.~~\cite{PSSB:PSSB2220480218}, the $E_B$ in the right hand side of Eq.~\eqref{repl} is replaced by the ``representative frequency'' which somewhat deviates from the binding energy.

The procedures outlined above have been effectively applied to excitons in bulk semiconductors~\cite{PSSB:PSSB2220850219,PSSB:PSSB2220480218,makarov:68} and have been recently applied to various TMDC monolayers~\cite{2018arXiv180106217S,Gao:2016aa,2018arXiv180401352S} with reasonable agreement with experimental data. However, as already mentioned, such an approach (and even direct numerical solution of Eq.~\eqref{BS1:1}) can be applied only if the frequency dispersion of $V^{\rm eff}_{q,\omega}$ is relatively weak: otherwise, omitted terms with self-crossings exemplified in Fig.~\ref{fig:BS}(c,d) can become important. It is worth stressing that the effects of the band renormalization and the state-filling due to the presence of charge carriers can be of importance, particularly, for excited states, cf. Ref.~\cite{2017arXiv170900891H}.

\section{Conclusions and outlook}

In general, the presence of free charge carriers, electrons or holes, changes the nature of the Coulomb interaction between the electron and hole forming an exciton.
It usually effectively reduces its strength as compared with the pristine situation where free charge carriers are absent. 
This screening effect depends on the spatial and temporal scales of the perturbation caused by the electron-hole interaction. The fact that the screening is typically effective at relatively large distances, i.e., for the wavevectors $q\lesssim q_s$ as further illustrated in Fig.\,\ref{fig:1}, is well established.
It needs to be accounted for both in conventional two-dimensional systems based on semiconductor quantum wells as well as in mono- and few-layer transition metal dichalcogenides and similar materials. 
By contrast, the role of the temporal dispersion of the screening, i.e., the retardation effect in the electron-hole interaction, generally remains a more involving and challenging issue. 
As one of the main consequences, it results in rather inefficient screening effect in TMDC monolayers under conditions, where the exciton binding energy is much higher in comparison to the Fermi energy of the free carriers.
In that case, the period of the electron-hole relative motion is so small that the free charge carriers are unable to accommodate to the rapidly changing electron-hole potential and thus do not efficiently screen the Coulomb interaction in the exciton. 

Quantitatively, the static screening is applicable provided that the exciton binding energy $E_B$ is smaller as compared to the electron (or hole) Fermi energy for degenerate charge carriers. 
For more typical Fermi energies $E_F \approx 0\ldots 50$~meV realized in both intentionally and unintentionally doped two-dimensional crystals the static screening model is shown to be largely inadequate.
The reasons are large binding energies of excitons on the order of several 100s of meV resulting from (i) relatively large effective masses and (ii) weak dielectric screening from the environment. 
Although for very high Fermi energies $E_F \gtrsim E_B$ the static approximation could be applicable, the effect of such high carrier densities on properties of the electronic states in TMDC monolayers is only little explored.
Moreover, such high densities may result in major qualitative changes of the band structure across the Brillouin zone.

The qualitative analysis presented here is supported by numerical calculations and comparison with available experimental observations.
In particular, the calculations of the binding energies and relative oscillator strengths of excitons demonstrate that the exciton resonances would vanish for typical densities of resident carriers in TMDC monolayers if the static screening concept were applicable. 
Moreover, it would also result in a very strong temperature dependence of these two quantities, which should converge to zero at cryogenic temperatures for typical carrier densities, including those due to unintentional doping in most cases. 
The main conclusions are further shown to be largely independent from the specific choice of the Coulomb potential. They are applicable both for the more traditional $1/{\rho}$ dependence and for the modified thin-film case, taking into account the non-uniform dielectric screening.

The predictions of the static approach are thus in stark contrast to the experimental results and theoretical considerations beyond the static screening regime. 
In fact, for reasonable free charge carrier densities the dynamical screening should indeed take place with the resulting efficiency being strongly diminished as compared with the static one. 
Overall, the extent to which the contributions from the static screening can be fully disregarded depends on the relation between the exciton binding energy and the electron (or hole) Fermi energy. 
Further quantitative and microscopic analysis thus remains important for the future experimental and theoretical studies.

Finally, it is interesting to note that similar considerations are applicable for the contributions of the lattice vibrations, i.e., phonons, to the screening of the Coulomb interaction in the exciton.
The dielectric polarization of the lattice follows the electric field produced by the electron and hole provided that the exciton binding energy is small compared to the polar phonon energies. 
The latter is often the case in many two-dimensional III-V and II-VI semiconductors, rendering the static dielectric constant generally applicable for the dielectric screening of excitons.
By contrast, for TMDC monolayers the exciton binding energies exceed by far the optical phonon energies both in the material itself and in the typical substrates or cap layers, including SiO$_2$ and hexagonal boron nitride. 
Although a semi-quantitative analysis of the excitonic states is potentially possible by using the high-frequency dielectric constants of the surrounding materials (in close analogy to the screening of the electron-hole exchange interaction where the transferred energy is even larger, on the order of the band gap~\cite{zhilich72:eng}), fully quantitative microscopic approaches remain of importance. 

Particularly interesting situation may arise if the energy of one of the phonon modes is in resonance with the binding energy or with the distance between the exciton levels. 
It further poses a question of whether the dielectric response of the material should be evaluated at different frequencies for the Rydberg series of exciton states with different binding energies.
In principle, it could mean that the binding energies of the individual levels are self-consistently determined not only by their spatial extent due to the dielectric contrast, as currently captured by the thin-film models, but also by the specific frequency dependence of the screening.
It could be thus instructive to address the transition from the high-frequency to a static screening regime for the exciton series with increasing principal quantum number (and thus decreasing binding energy). 
In this regard, the ideas and the experiments similar to the recent reports such as Ref.~\cite{Klots:2018aa} are of particular importance.

Similar considerations may apply for more complex excitonic states such as trions and biexcitons.
While the microscopic nature of these quasiparticles is currently under intense debate, particularly, due to the many-electron effects, the general question of the appropriate frequency range to evaluate dielectric screening remains.
In this respect, it is not obviously clear that the binding energies of trions and biexcitons, defined as the energy distance between the peaks in optical response of the trion/biexciton and exciton, provide an equally valid scale for comparison with the excitations of the screening medium.
Alternatively, the average kinetic energy of the individual trion and biexciton constituents could be a more reasonable parameter to consider, which is rather similar to the above discussed scenario for neutral excitons.   
We therefore note, that while a more involved analysis of this particular topic is beyond the scope of the present study, it certainly deserves more attention in the future research. 
Theoretically, it requires full treatment of the manybody system. Experimentally, the detailed studies analogous to those reported in Ref.~\cite{Klots:2018aa} are highly anticipated.

On the theoretical side, both analytical and numerical calculations with retarded electron-hole interaction along the lines presented above seem to be required in the case of the free carrier and, potentially, also lattice phonon screening.
With respect to that, alternative approaches such as invoking the polaron concept, including both exciton-phonon- and Fermi-polaron\,\cite{suris:correlation,PhysRevB.95.035417,Sidler:2016aa,PhysRevB.95.035417} could be particularly important.
Then the ansatz for the solution of many-body problem of either ``exciton+phonon bath'' or ``exciton+Fermi sea'' would, in principle, allow one to account for the correlations between the electron-hole pair and the many-body bath excitations.
Complementary, microscopic approaches based on semiconductor-Bloch equations and cluster expansion\,\cite{Steinhoff2014,Schmidt2016,Meckbach2018}, are highly viable in that respect as well.

\section{Acknowledgments}

We thank Timothy C. Berkelbach, David R. Reichmann, Diana Qiu, Ting Cao, Tim Wehling, Ermin Malic, Andreas Knorr and Tony F. Heinz for interesting and helpful discussions.
Financial support by the German Science Foundation (DFG) via Emmy Noether Grant CH 1672/1-1 and Collaborative Research Center SFB 1277 (B05) is gratefully acknowledged. 
M.M.G. is grateful to the Russian Science Foundation (Grant No. 14-12-01067).


%

\end{document}